# Superconductivity in an Orbital-reoriented SnAs Square Lattice: a Case Study of Li$_{0.6}$Sn$_2$As$_2$ and NaSnAs

Junjie Wang†[a], Tianping Ying*†[a], Jun Deng†[a], Cuiying Pei†[b], Tongxu Yu[c], Xu Chen[a], Yimin Wan[d], Mingzhang Yang[a], Weiyi Dai[c], Dongliang Yang[e], Yanchun Li[e], Shiyan Li[d], Soshi Iimura[f], Shixuan Du[a], Hideo Hosono[f], Yanpeng Qi*[b],　Jian-gang Guo*[a,g]

[a] J.J. Wang, Dr. J. Deng, Dr. X. Chen, M. Z. Yang, Prof. Dr. S. X. Du, Prof. Dr. T. P. Ying, Prof. Dr. J.-G. Guo
Institute of Physics and University of Chinese Academy of Sciences, Chinese Academy of Sciences, Beijing 100190, China
E-mail: ying@iphy.ac.cn; jgguo@iphy.ac.cn

[b] Dr. C. Y. Pei, Prof. Dr. Y. P. Qi
School of Physical Science and Technology, ShanghaiTech University, Shanghai 201210, China
ShanghaiTech Laboratory for Topological Physics, ShanghaiTech University, Shanghai 201210, China
Shanghai Key Laboratory of High-resolution Electron Microscopy, ShanghaiTech University, Shanghai 201210, China
E-mail: qiyp@shanghaitech.edu.cn

[c] Dr. T. X. Yu, Dr. W. Y. Dai
Gusu Laboratory of Materials, Jiangsu 215123, China
Suzhou Laboratory, Jiangsu 215123, China

[d] Y. M. Wan, Prof. Dr. S. Y. Li, State Key Laboratory of Surface Physics and Department of Physics, Department of Physics, Fudan University, Shanghai 200438, China

[e] Dr. D. L. Yang, Prof. Dr. Y. C. Li
Beijing Synchrotron Radiation Facility and Institute of High Energy Physics, Chinese Academy of Sciences, Beijing, 100049, China

[f] Prof. Dr. S. Iimura, Prof. Dr. H. Hosono
National Institute for Materials Science (NIMS), Tsukuba, Ibaraki, 305-0047, Japan
Materials Research Center for Element Strategy, Tokyo Institute of Technology, Yokohama, 226-8503, Japan

[g] Prof. Dr. J.-G. Guo
Songshan Lake Materials Laboratory, Dongguan, Guangdong 523808, China

† These authors contribute equally.

**Abstract:** Searching for functional square lattices in layered superconductor systems offers an explicit clue to modify the electron behavior and find exotic properties. The trigonal $SnAs_3$ structural units in SnAs-based systems are relatively conformable to distortion, which provides the possibility to achieve structurally topological transformation and higher superconducting transition temperatures. In the present work, the functional As square lattice was realized and activated in $Li_{0.6}Sn_2As_2$ and NaSnAs through a topotactic structural transformation of trigonal $SnAs_3$ to square $SnAs_4$ under pressure, resulting in a record-high $T_c$ among all synthesized SnAs-based compounds. Meanwhile, the conductive channel transfers from the out-of-plane $p_z$ orbital to the in-plane $p_x+p_y$ orbitals, facilitating electron hopping within the square 2D lattice and boosting the superconductivity. The reorientation of *p*-orbital following a directed local structure transformation provides an effective strategy to modify layered superconductors.

**Introduction**

Attributing the emergent properties of compounds to structural units is naturally a reductionism standpoint, but is highly desirable and easily operable to understand, manipulate, and predict the performance of the systems under various physical conditions. In this aspect, the well-defined square lattice plane in high transition temperature ($T_c$) superconductors, e.g., cuprates[1],[2] and iron-based[3],[4] systems is deemed as an indispensable ingredient and plays a unique role in penetrating the puzzle of underlying pairing mechanisms. The cuprates share the square Cu-O plane where the electrons propagate through the Cu *d*-orbitals with the super-exchange interaction of oxygen *p*-orbitals[5],[6],[7]. Adding an apical oxygen to enhance the out-of-plane dispersion of the electron will localize the in-plane electronic wave function, and ultimately lower the maximum $T_c$[8]. Similarly, the key ingredient of iron-based superconductors is the spatial repetition of FeAs/FeSe tetrahedron units to form a layered structure with alternating square lattices of iron and As/Se planes[9],[10],[11],[12],[13], in which the distortion of the FeSe from planar-tetragonal to 3D-trigonal connections under high pressure is accompanied by the complete loss of superconductivity (SC)[14]. From a stereochemistry perspective, the square tiling can make full use of the orthogonal *d*- and *p*-orbitals, enabling easy charge carrier hopping between adjacent lobes[15]. Thus, searching novel layered systems with square motifs is a promising strategy for exploring high-$T_c$ superconductors.

SnAs-based compounds are such auspicious candidates, where the $T_c$s of layered compounds are surprisingly lower than their 3D counterparts. Layered SnAs-based compounds have diverse combinations with alkali[16],[17],[18],[19], alkaline-earth[20], rare-earth metals[21], and even molecule clusters[22], resulting in highly tunable carrier concentrations and stacking sequence. Despite the structural flexibility, however, they are often semiconducting or barely superconducting at low temperatures of 1.1~1.3 K[16],[17],[18],[19],[20],[21],[22]. The binary SnAs with a rock-salt structure ($Fm$-3m), in contrast, has a relatively higher $T_c$ of 4 K[23],[24]. A naive interpretation of the lowered $T_c$ in layered-SnAs compounds is related to the trigonal pyramid $SnAs_3$ unit. It is the steric effect of the misaligned $p$-orbitals that somehow impedes electron flow within the superconducting layer.

High pressure can create atypical materials by repopulating atomic orbitals and reconstructing functional units without altering the chemical composition[25],[26],[27],[28]. Herein, we explore the possibility of obtaining higher-$T_c$ superconductor by constructing square-lattice configurations under pressure, employing layered metallic $Li_{0.6}Sn_2As_2$ and semiconducting $NaSnAs$ as examples. Theoretical calculations predict a topotactic transformation of the functional units from $SnAs_3$ to $SnAs_4$, accompanied by a trigonal-tetragonal phase transition, which is verified by our *in-situ* synchrotron diffraction measurements. Prominently, the tetragonal $Li_{0.6}Sn_2As_2$ shows the highest $T_c$ of 7.5 K among all the synthesized SnAs-based compounds that is six times higher than its ambient value. Pressurized NaSnAs exhibits a similar $SnAs_3$-$SnAs_4$ transformation and enhanced SC, but with a more complex three-stage phase transformation from semiconductor to two successive superconducting phases. Theoretical analyses of their orbital components reveal an intriguing charge redistribution from out-of-plane to in-plane, which enhances electron hopping and contributes to the record high $T_c$ in the SnAs square lattice.

**Results and Discussion**

Ambient structures ($\alpha$ phase) of $Li_{0.6}Sn_2As_2$ ($R\bar{3}m$, Fig. 1a) and NaSnAs ($P6_3mc$, Fig. S1) share the similar trigonal pyramid of $SnAs_3$ building blocks, with the intercalation of alkali metals in between one or two SnAs layers. As shown in Fig. 1c, the electrons from $5p$ of Sn, $4p$ of As and one extra electron from the $Li^+/Na^+$ construct the bonding state of $SnAs_3$, leaving a lone pair from $5s^2$ dangling along the $c$ axis, away from the $SnAs_3$ pyramid to create the canonical lone-pair electrons[26],[29],[30]. Due to the fully occupied Sn-As bonds and the repulsive interaction of lone-pair electrons, NaSnAs turns out to be a semiconductor with a band gap of 0.31 eV[31]. It is therefore easy to understand the metallic nature of $Li_{0.6}Sn_2As_2$ (0.3 e/SnAs) and $NaSn_2As_2$ (0.5 e/SnAs) with reduced electron doping, which further become superconducting with similar $T_c$ of 1.5 K (Note that $Li_{0.6}Sn_2As_2$ is a newly discovered superconductor, and its properties are shown in Figs. S2 and S3). Given that the electrons move within the SnAs plane in a zigzag manner (illustrated as dashed curve in Fig. 1c), it is possible to alter the $SnAs_3$ configuration by pressure, which may facilitate the flow of electrons. Therefore, we begin with determining the structure under pressure through theoretical structure search package CALYPSO[32],[33]. During the structure searching, the size of unit cells is limited from 2 up to 4 formulas for each stoichiometry (see Fig. S4). Both the generation sizes and the number of generations were set to 30 for getting a converged result.

Our structural searches up to 30 GPa reveal that a new phase with a space group of *I*4/*mmm* (β phase) is the most stable one. We superimposed the β phase on the α phase as open circles to illustrate the structural transformation (Fig. 1b and Fig. S5). The SnAs$_3$ building blocks only need to bear a mild distortion under this transformation. This is reasonable if considering that the coordination numbers generally increase under high pressure to homogenize the electron distribution [34],[35]. We check the pressure-dependent enthalpy of the β phase with respect to that of the α phase from 0 to 50 GPa in Fig. 1(d). Beyond a critical pressure of around 25 GPa, the β phase overwhelms the α phase in energy. Figure S6 shows the phonon dispersion of the β phase at 30 GPa, which does not have any imaginary frequencies, indicating the stability of the high-pressure phase. It is worth noting that the tetragonal pyramid configuration of SnAs$_4$ has not been previously reported.

We also did the structural search of NaSnAs under pressure. At low pressure, NaSnAs undergoes a structural transformation from the α phase to the β phase at 15 GPa, with the lowest enthalpy in the Na-Sn-As system. However, calculations at 30 GPa predict that another tetragonal phase with the space group *P*4/*mmm* (γ phase) is the most stable. Figure S7 depicts the successive structural transformation of the three phases. The γ phase can be viewed as a change in the stacking sequence of the Na spacer layer from the β phase, requiring every other Na layer transmits through its adjacent SnAs layers. The β phase is predicted to be stable in a pressure window of 15-20 GPa, as indicated by the pressure-dependent formation enthalpy shown in Fig. 1e. Obviously, the interlayer transmission of Na from the β phase to γ phase should be more energetically costly than the intralayer SnAs$_3$ to SnAs$_4$ transformation. Nonetheless, both β phase and γ phase share the identical SnAs$_4$ building block.

To verify the SnAs$_3$-SnAs$_4$ transformation, we performed *in-situ* synchrotron diffraction experiments on Li$_{0.6}$Sn$_2$As$_2$ at various pressures ranging from ambient pressure to 48.0 GPa. All diffraction peaks show a systematic shift below 25.0 GPa and can be indexed into the α phase (Fig. S8). An additional peak at 15.5° gradually arises as pressure above 25.0 GPa, signaling a pressure-driven phase transition. When the pressure is increased to 40 GPa, the initial peaks corresponding to the α phase become almost indistinguishable. We carried out Rietveld refinements on each synchrotron x-ray diffraction pattern, and put three typical profiles of 48.0 GPa, 32.8 GPa, and 1.88 GPa in Fig. 2a. We refined the pattern of 48.0 GPa based on the theoretically predicted β phase and obtained the agreement factors of $R_p$ = 0.84% and $R_{wp}$ = 5.3%, which are comparable to the ambient refinement values of $R_p$ = 1.21% and $R_{wp}$ = 5.43%, respectively. A two-phase refinement against the pattern of 32.8 GPa produces the best fitting result. We note that the diffraction pattern depressurized phase is almost identical to the initial one except mild peak broadening (Fig. S9), indicating the reversibility of the SnAs$_3$-SnAs$_4$ transformation.

The contour plot of the diffraction patterns reveals more details of the structural transition (Fig. 2b). The critical pressure of 25.0 GPa is clearly with the emergence of new diffraction peaks at 15.5° and 20.2°. An interesting observation is that the diffraction peaks corresponding to the α phase (102, 105, 110, 025, and 207 in particular) exhibit a consistent kink at 25.0 GPa, indicating substantial lattice distortion in the α phase as an intermediate transition state. A qualitative analysis of the volume fractions of α and β phase extracted from Rietveld refinements are shown in Fig. 2c.

Another notable thing is the dramatic change in bond lengths under pressure. Figure S10 labels the Sn-As and As-As distances at 0 and 30 GPa. The out-of-plane As-As distance is reduced from 4.04 Å to 2.58 Å and the in-plane As-As distance shrinks abruptly from 4.01 Å to 3.51 Å along with the $SnAs_3$-$SnAs_4$ transformation. However, the Sn-As bond is increased to 2.97 Å, which is even larger than the ambient value of 2.72 Å. The elongation of the Sn-As bond suggests that the prior conduction channel through As-Sn-As is greatly impeded, whereas direct hopping between As-As anions within the square lattice should be much favored.

The increased coordination numbers, elongated Sn-As bonds, and collapsed As-As distance induced by the $SnAs_3$-$SnAs_4$ transformation significantly affect the electronic transport properties. The α-$Li_{0.6}Sn_2As_2$ exhibits only one superconducting transition below 27.7 GPa, with its $T_c$ gradually increasing from 1.3 K to 5.3 K (blue arrows in Fig.3a). Noticeably, a drop in resistivity at 7.5 K is observed at 27.7 GPa (enlarged rectangle shown in Fig. 3a), which is consistent with the emergence of the β phase around 25.0 GPa. To the best of our knowledge, this is the highest $T_c$ experimentally realized in the SnAs-based compounds, six times higher than the ambient value. As pressure increasing, the decline in resistivity gets more pronounced, yet its transition temperature remains nearly constant (red arrows in Fig. 3a). All the raw data are shown in Fig. S11. We confirm that this newly discovered kink in resistivity is superconductive according to the gradual suppression of the $T_c$ under magnetic fields (Fig. 3c). The superconducting drop corresponding to the α phase is seen at 53.1 GPa, indicating the survival of the trace α phase. This is also in line with our estimation of the volume fraction shown in Fig. 2c.

Figure 3b reveals the $T_c$ evolutions of the α phase and β phase in response to pressure. The $T_c$ in the β phase remains nearly constant at moderate pressure (25.0~45.0 GPa) and then decreases somewhat. Their $T_c$s become to be one transition above 53.0 GPa. Given that the structure at high pressure is dominated by the β phase, the remaining superconducting transition should be attributed to the $SnAs_4$ arrangement. Note that the maximum $T_c$ is achieved in the mixture phase rather than the pure β phase. A plausible reason for this could be that the $T_c$-pressure diagram of the β phase has a dome-like shape, within which the maximum $T_c$ occurring ~35 GPa. Figure 3d shows the $H_{c2}(0)$ of α (0 GPa) and β (73 GPa) phases. Linear extrapolation shows a higher $H_{c2}(0)$ in the β phase.

The ensuing transition from α to β to γ phases in NaSnAs is investigated by transport measurements. NaSnAs undergoes a semiconductor-metal transition under lower pressure, then becomes SC at ~10 GPa (shown in Fig. S12). Its $T_c$ slowly increases to 3.6 K at 35 GPa, then suddenly jumps to 6.8 K. While the onset pressure of the first superconducting phase at ~10 GPa is qualitatively consistent with the theoretical pressure of α-β phase transition, the pressure of $T_c$'s jump is higher than our theoretical prediction of 20 GPa (Fig. 1e). This is understandable because the predicted γ phase is a thermally equivalent ground state. Since our *in-situ* high-pressure measurements are carried out in a DAC cell at room temperature, the energy barrier for the inter-plane migration (β-γ) should be much higher than the intra-plane ion displacement (α-β). Despite their different carrier doping levels, the charge-balanced NaSnAs achieves a comparable $T_c$ to that of the β-$Li_{0.6}Sn_2As_2$, indicating the highest $T_c$ in the SnAs system is determined the common building block of $SnAs_4$ in terms of structure.

The structural transformation of $SnAs_3$-$SnAs_4$ can lead to charge rearrangements of conducting electrons. As shown in Fig. S13, the Fermi level ($E_F$) in $LiSn_2As_2$ is mostly composed of As's $4p$ and Sn's $5p$ orbitals, with the $4s$ and $5s$ orbitals lying far below the $E_F$. We plot the orbital projected density of states (PDOS) in Fig. 4a. At ambient pressure, the PDOS of As's $p_z$ component at 0.98 states/eV is higher than the sum of its $p_x$ and $p_y$ orbital (0.76 states/eV). This feature is more prominent for Sn because the intensity ratio of $p_z/(p_x+p_y)$ reaches 3.6 at the $E_F$. Specifically, the spectra weights of $p_z$ and $p_x+p_y$ are reversed for both As and Sn at high pressure. Specifically, the enhancement of the PDOS of As's $p_x+p_y$ is 4 times higher than that of the $p_z$ component. The $SnAs_3$-$SnAs_4$ transformation induces a prominent charge redistribution from the out-of-plane ($p_z$) orbital to the in-plane ($p_x$, $p_y$) ones. We perform Crystal Orbital Hamilton Population (COHP) calculations for the α and β phases. As shown in Fig. 4b, the α phase is dominated by Sn-As bonding states below -1 eV, with a small component at the $E_F$. In the β phase, however, the out-of-plane As-As bonding component of inter $SnAs_4$ units overwhelms that of Sn-As bonding of intra unit, and dominates the $E_F$ by As-As anti-bonding. Moreover, the integrals of COHP (ICOHP) up to the $E_F$ for Sn-As and As-As are -3.16 and -0.08 eV/pair in the α phase, and change to -1.48 and -3.89 eV/pair in the β phase, respectively. This suggests that the interaction of Sn-As dominates in the α phase, while that of As-As prevails in the β phase.

We plot the partial electron distribution near the $E_F$ (-0.2~0.2 eV) to see the charge distribution. For the α-$LiSn_2As_2$, the valence electrons mainly gather around Sn and As atoms and are oriented along the *c* axis (Fig. 4c, left). Thus the electrons can mainly migrate along As-Sn-As via zigzagging through the shared $5p_z$-$4p_z$ orbitals (Fig. 4d left panel and Fig. 1c right panel). Once the $SnAs_3$ is transformed into $SnAs_4$, the charge redistributions from $p_z$ to $p_x+p_y$ are clearly seen in the right panels of Fig. 4c and 4d. We examined the partial electron distribution of the β-$LiSn_2As_2$ in each energy sector. With the electron filling from the lowest occupied σ bond to the σ* bond, their respective electron dispersion in real space is shown in Fig. S14.

Summarizing all the analyses above, we provide a simple molecular orbital diagram to understand the evolution of bond connections in LiSn$_2$As$_2$ (Fig. 4f and 4g). The out-of-plane As-As interaction drives the $E_F$ sitting on the $\pi_{xy}^*$ band with the extra electron donation from Li and Sn, which well explains the calculated dominating $p_x+p_y$ PDOS (lower panel in Fig. 4a) and the As-As anti-bonding (lower panel in Fig. 4b). Considering the orthogonal configuration of $p_x$ and $p_y$ orbitals, the electron is more likely to migrate within this As-square lattice than in the triangle ones (Fig. 4e). The abrupt shrinkage in the As-As distance from 4 Å to 3.5 Å, on the other hand, reinforces in-plane hopping via overlapping of the orthogonal $p_x+p_y$ lobes. This reminds us of the highly tunable Bi-Bi distance from 3.85 to 4.08 Å in Bi-based square lattice $R_2O_2$Bi ($R$ = rare earth metals), accompanied by a metal-insulator transition at a shortened Bi-Bi distance of 3.95 Å in Sm$_2$O$_2$Bi[36] and a subsequent SC at 3.87 Å in Y$_2$O$_2$Bi[37].

As our simulations are all based on the fully occupation of Li in LiSn$_2$As$_2$, the influence of Li vacancies is carefully checked. To this end, we create a 3×3×1 superlattice and randomly removed 7 out of 18 Li atoms in the supercell corresponding to Li$_{0.61}$Sn$_2$As$_2$. Figure S15 shows the partial charge density at 30 GPa near $E_F$ (-0.2~0.2 eV) of the model. It again shows a $p_x/p_y$ conducting character, indicating that the Li vacancy does not influence our main result.

The electronic states in the pressurized NaSnAs resemble those of Li$_{0.6}$Sn$_2$As$_2$. As shown in Fig. S16, the $E_F$ of both β and γ phases are dominated by $p_x+p_y$ orbitals. The 2D feature of the electron dispersion becomes more prominent in the γ phase (Fig. S17). It is the charge transfer from $p_z$ to $p_x+p_y$ that facilitates the direct hopping of electrons. Given the similarity of charge distribution in the SnAs$_4$ unit, we conclude that the orbital reorientation of As's $p$ orbitals in a square lattice should be responsible for the similar $T_c$ in both compounds.

**Conclusion**

In this work, a new kind of structure unit SnAs$_4$ is realized in SnAs-based compounds through a topotactic phase transformation in both Li$_{0.6}$Sn$_2$As$_2$ and NaSnAs. Thanks to the thus-formed As square lattice and the dramatic shrinkage of the in-plane As-As distance from SnAs$_3$ to SnAs$_4$ transformation, the probability of in-plane hopping through $p_x+p_y$ lobes is much enhanced. Theoretical calculations show a charge redistribution of the electron near the $E_F$ from $p_z$ orbital to $p_x+p_y$ orbitals, further facilitating the electron transportation and the subsequent record-high $T_c$ in all reported SnAs-based compounds. Our findings verify the effectiveness of modulating SC through directed local structure tailoring of flexible layered compounds.

**Acknowledgements**

We appreciate Prof. Xianxin Wu for fruitful discussion. This work is financially supported by the National Key Research and Development Program of China

(No. 2018YFE0202600, 2021YFA1401800), Beijing Natural Science Foundation (Grant No. Z200005), the National Natural Science Foundation of China (No. 51922105, 11804184, and 11974208), the Strategic Priority Research Program of the Chinese Academy of Sciences (Grant XDB30000000). ADXRD measurements were performed at 4W2 High Pressure Station, Beijing Synchrotron Radiation Facility (BSRF), which is supported by Chinese Academy of Sciences (Grant KJCX2-SW-N20, KJCX2-SW-N03). Y.Q. would like to acknowledge the National Key R&D Program of China (Grant No. 2018YFA0704300), the National Natural Science Foundation of China (grant nos. U1932217, 11974246, and 12004252). The authors thank the support from CℏEM (02161943) and Analytical Instrumentation Center (SPST-AIC10112914), SPST, ShanghaiTech University. H. H. was supported by a grant from the MEXT Element Strategy Initiative to Form Core Research Center (No. JPMXP0112101001) and JSPS Kakenhi Grants-in-Aid (No. 17H06153).

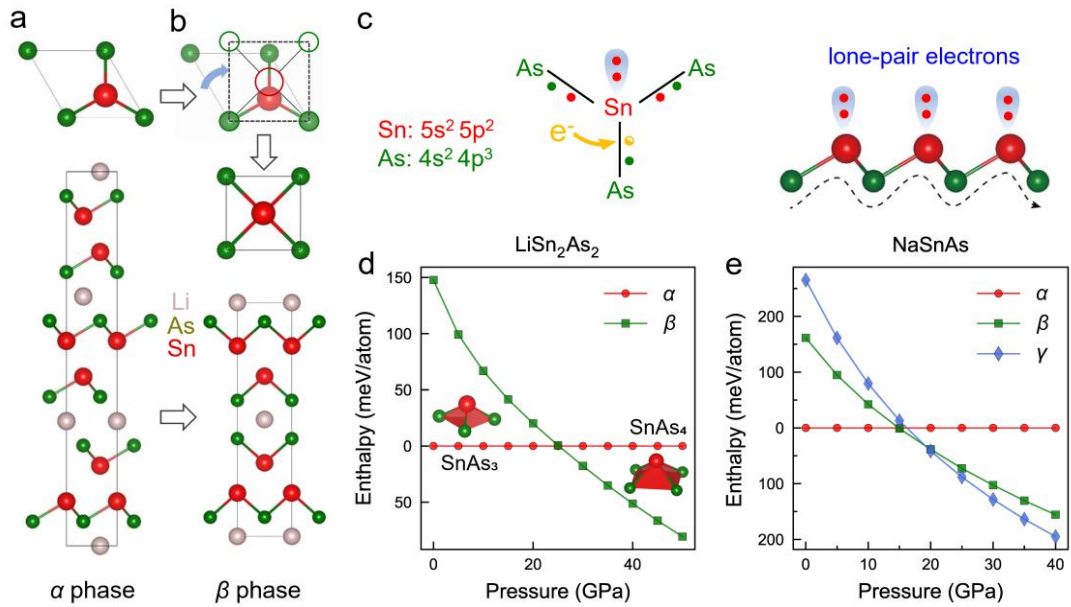

**Figure 1**. Theoretical prediction of the structure transition of LiSn$_2$As$_2$ and NaSnAs. Top and side-views of LiSn$_2$As$_2$ at a) ambient pressure ($R\bar{3}m$, α phase) and b) high pressure ($I4/mmm$, β phase) predicted by the CALYPSO package. c) Bonding environment of SnAs-layered compounds and the illustration of the zigzag-propagation of electrons within the *ab* plane. The lone-pair electrons are denoted by light blue droplets. d) Relative enthalpy of LiSn$_2$As$_2$ as a function of pressure. The insets depict the trigonal (SnAs$_3$) and tetragonal (SnAs$_4$) pyramids. e) Relative enthalpy of $P6_3mc$ (α phase), $I4/mmm$ (β phase) and $P4/mmm$ (γ phase) for NaSnAs as a function of pressure. The enthalpy of the α phase is taken as a reference.

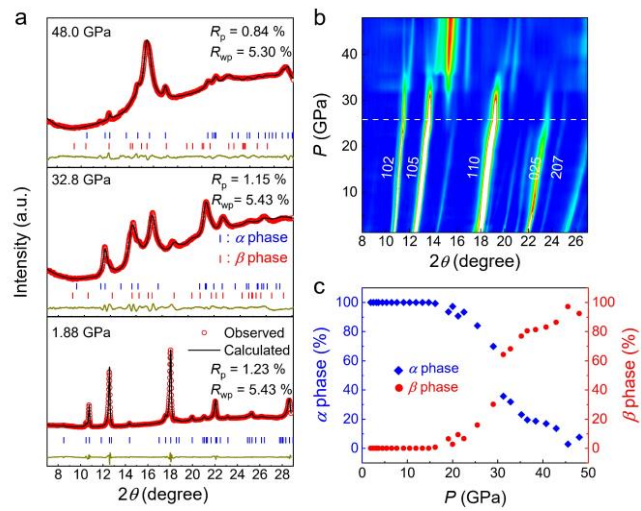

**Figure 2.** *In-situ* synchrotron diffractions of Li$_{0.6}$Sn$_2$As$_2$ at pressures. a) Three Rietveld refinements profiles of Li$_{0.6}$Sn$_2$As$_2$ at 48.0 GPa, 32.8 GPa and 1.88 GPa. b) Color contour plot of the pressure-dependent diffraction peaks from 1.88 to 48.0 GPa. c) Pressure-mediated volume fraction of the *α* phase and *β* phase extracted from refinement.

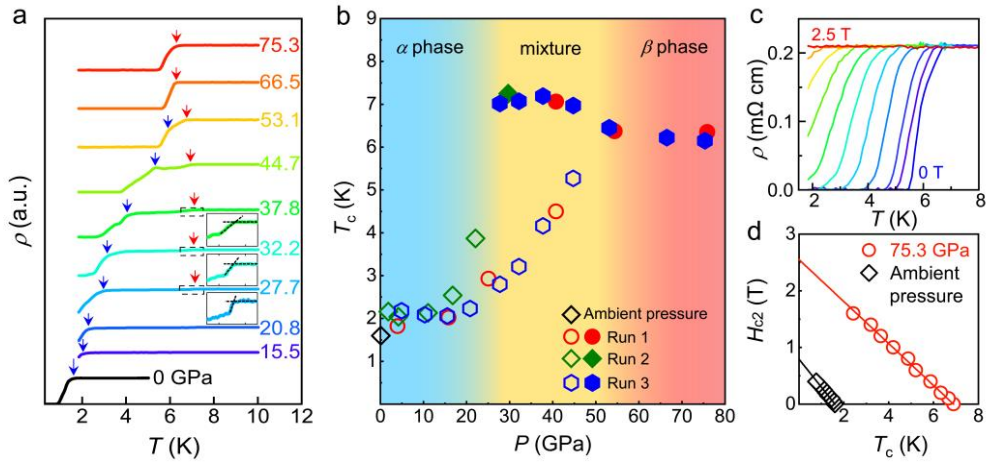

**Figure 3**. Pressure-dependent SC in Li$_{0.6}$Sn$_2$As$_2$. a) Temperature-dependent resistivity of Li$_{0.6}$Sn$_2$As$_2$ under pressure. Insets are the enlargement of resistivity at higher temperatures. Blue and red arrows indicate the onset $T_c$ of $α$ and $β$ phases. b) Superconducting phase diagram of Li$_{0.6}$Sn$_2$As$_2$ under pressure. An intermediate pressure region mixes two transition temperatures. c) Resistivity of the $β$ phase (75.3 GPa) from 1.5 K-8 K at different magnetic fields. d) Fitting of the upper critical fields for the ambient (black squares) and high pressure (red circles) phases.

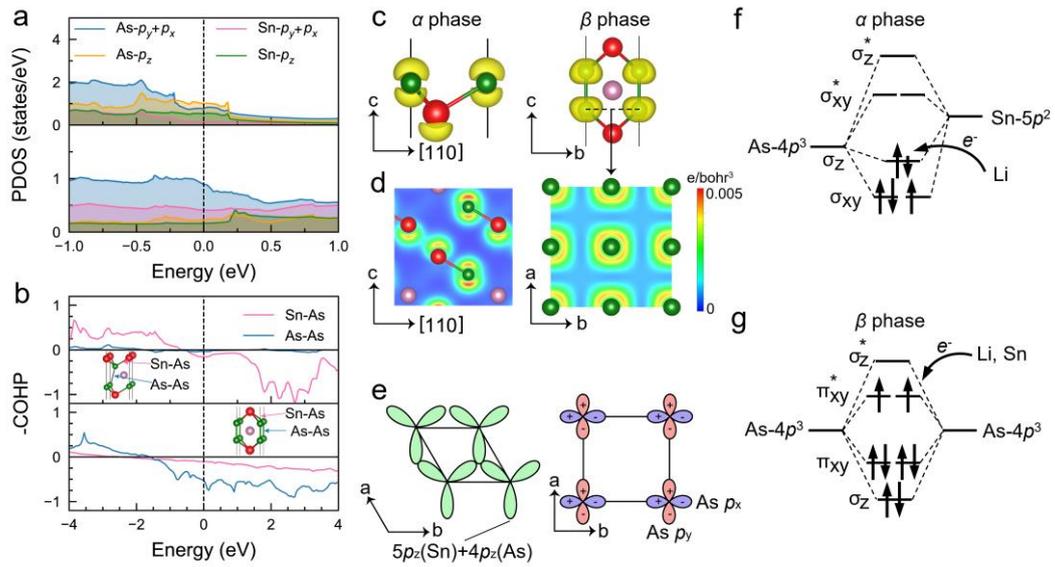

**Figure 4**. Charge redistribution and orbital reorientation from SnAs$_3$-SnAs$_4$ transformation in LiSn$_2$As$_2$. a) Projected density of states (PDOS) and b) -COHP for *α* phase at 0 GPa (upper panel) and *β* phase at 30 GPa (lower panel), respectively. Insets in b) show the interaction of As-As and Sn-As. Positive values indicate bonding states, and negative ones are antibonding states. c-e) Partial charge density around the $E_F$ (-0.2 eV ~0.2 eV) with iso-value 0.0025 e/Bohr$^3$, section projections, and illustration of bond connections of the *α* phase (left) and *β* phase (right), respectively. f-g) Molecular orbital diagram of the Sn-As bonding state in the *α* phase and out-of-plane As-As antibonding state in the *β* phase. Note that extra electrons in the π$_{xy}$* band come from the donation of Li and Sn.

13